\documentclass[twocolumn,showpacs,superscriptaddress]{revtex4}  

\usepackage[english]{babel}
\usepackage[latin1]{inputenc}

\usepackage{dcolumn}
\usepackage{subfigure}
\usepackage{multirow}
\usepackage{amsmath,amsfonts,amssymb}

\usepackage{graphicx}  
\usepackage{latexsym}   

\begin{document}

\title{Electromagnetic energy within a magnetic infinite cylinder and scattering properties for oblique incidence}

\author{\firstname{Tiago}  Jos\'e  \surname{Arruda}}
\author{\firstname{Alexandre} Souto \surname{Martinez}}
\email{asmartinez@ffclrp.usp.br}

\affiliation{Faculdade de Filosofia,~Ci\^encias e Letras de Ribeir\~ao Preto,\\
             Universidade de S\~ao Paulo \\
             National Institute of Science and Technology in Complex Systems \\
             Avenida Bandeirantes, 3900 \\
             14040-901, Ribeir\~ao Preto, S\~ao Paulo, Brazil.}

%
%

\begin{abstract}

In this work we analytically calculate the time-averaged electromagnetic energy stored inside a nondispersive magnetic isotropic cylinder which is obliquely irradiated by an electromagnetic plane wave.
An expression for the optical-absorption efficiency in terms of the magnetic internal coefficients is also obtained.
In the low absorption limit, we derive a relation between the normalized internal energy and the optical-absorption efficiency which is not affected by the magnetism and the incidence angle.
This mentioned relation, indeed, seems to be independent on the shape of the scatterer.
This universal aspect of the internal energy is connected to the transport velocity and consequently to the diffusion coefficient in the multiple scattering regime.
Magnetism favors high internal energy for low size parameter cylinders, which leads to a low diffusion coefficient for electromagnetic propagation in 2D random media.

\end{abstract}


\maketitle


\section{Introduction}

The interest in the study of magnetic materials and their optical properties has been increased in the applied science.
Recently, applications on magnetic 2D and 3D photonic band gaps (PBGs) \cite{sigalas,ping,liu,lin2}, microwave filters, metamaterials \cite{lin}, high density magnetic recording media \cite{tarento}, weak localization of light \cite{alexandre-vanish,alexandre-multi} have been reported.
All these works explore the fact that, at microwave or radio frequencies, the magnetic materials exhibit large values of magnetic permeability \cite{sigalas,ping}.

Electromagnetic (EM) scattering by magnetic spheres has been applied by Kerker et al.~\cite{kerkermag}.
Some unusual features present in single and multiple magnetic Mie scattering, such as forward-backward asymmetry with preferential backward scattering and resonance effects \cite{alexandre-new,alexandre-elec}, and vanishing of the energy-transport velocity even for small size parameters \cite{alexandre-vanish,alexandre-multi}, have been studied.
In a recent paper, we have calculated the EM energy stored inside a magnetic sphere and we have shown that, even for size parameters much smaller than unity (Rayleigh size region), this quantity is strongly enhanced, with sharp resonance peaks \cite{tiago}.

The problem of EM scattering by an isotropic circular cylinder is not new \cite{rayleigh,stratton}.
For a general case of oblique incidence and magnetic scatterers, an analytical solution has been provided a long time ago by Wait \cite{wait} and has also been treated by Lind and Greenberg \cite{lind} in the context of dielectric infinite cylinders.
Although expressions for the stored energy in a normally illuminated dispersive and nondispersive dielectric cylinders have been documented in the literature \cite{ruppin}, no analytical study of the magnetism influence at a general case of oblique incidence has been so far performed.
Our aim is to fill this gap with a detailed study of a cylindrical magnetic scatterer illuminated at an arbitrary incidence angle.
We devote special attention to the fields inside the scattering center and their application on the calculation of the energy-transport velocity in a disordered magnetic medium \cite{ruppin,bart}.

The framework in which the scattering quantities are calculated is presented in Sec.~\ref{theory} of this description.
Essentially, we present the main expressions obtained solving the macroscopic Maxwell's equations for the EM internal fields \cite{wait,bohren}.
We have adopted the same notation as Bohren and Huffman \cite{bohren}.
In Sec.~\ref{internal}, for a general case of oblique incidence, we use new relations among Bessel function to calculate the scattering coefficients and consequently the normalized average EM energy stored inside a magnetic infinitely long cylinder.
This extends the study of \cite{ruppin,bott} for nondispersive scatterers.
Also, in Sec.~\ref{absorption}, we derive exact and approximated expressions for the optical-absorption efficiency in terms of the internal coefficients.
This last result, which has an analogue in the single Mie scattering \cite{tiago}, is important to link measurable quantities with the time-averaged EM energy \cite{bott,tiago}.
Numerical results are shown in Sec.~\ref{numerical}.
Specially, for a weakly absorptive magnetic cylinder, we determine a relation between the internal energy-enhancement factor and the absorption efficiency, which does not depend on the incidence angle and the polarization of the incident EM wave.
Comparing our result with that one obtained firstly by \cite{bott}, we ascribe the achieved difference only to the cylindrical and spherical geometries.
This geometrical consideration allows us to write a relation which is independent on the shape of the scatterer.
Following \cite{ruppin,bart}, we present an application of our calculations to the study of the energy-transport velocity.
For a two-dimensional disordered magnetic medium, we obtain an oscillatory behavior of the energy-transport velocity, as a function of the size parameter, even in the Rayleigh size region.
Briefly, in Appendix~\ref{appendice}, we present approximations for the far-field scattering coefficients and some considerations about the degree of polarization for magnetic cylinders at normal incidence.

\section{Basic theory}
\label{theory}

Let the scatterer be an infinite right circular cylinder with finite radius $a$ embedded in an infinite non-absorptive medium.
Both the cylinder and the surrounding medium are assumed to be linear, homogeneous and isotropic, with inductive capacities ($\epsilon_1,\mu_1$) and ($\epsilon,\mu$), respectively.
The incident EM wave, which interacts with the cylinder, is a plane and monochromatic complex wave, with time-harmonic dependence given by $\exp({-\imath \omega t})$ \cite{bohren,barber}.
The quantity $\omega$ is the angular frequency and it is considered to be the same for the incident and scattered waves (elastic scattering).
In addition, suppose that these media are electromagnetically source-free and adopt the international system of units.
On account of symmetry, the cylindrical scatterer imposes two basic linear polarizations for the incident EM wave \cite{bohren,hulst}.
They are referred to as the TM (or case I) and TE (or case II) modes \cite{bohren}.
In the former, the incident electric field is parallel to the $xz$ plane, while in the latter it is perpendicular to this plane \cite{bohren,hulst,barber,kerker}.
For both cases, consider that $\zeta$ is the angle between the wavevector $\mathbf{k}$ and the $z$ axis, where $k=|\mathbf{k}|=\omega({\mu\epsilon})^{1/2}$ is the wavenumber.

Inside the cylinder ($0\leq r\leq a$), the expansion of the internal EM field ($\mathbf{E}_1,\mathbf{H}_1$) in terms of cylindrical harmonics $\mathbf{M}_n$ and $\mathbf{N}_n$ \cite{bohren} are expressed below.
For the TM and TE modes, which are indicated by the indexes (I) and (II), respectively, one obtains
\begin{eqnarray}
    \mathbf{E}_1^{\rm (I)}&=&\sum_{n=-\infty}^{\infty}E_n\left[d_n^{\rm(I)}\mathbf{M}_n^{(1)}+c_n^{\rm(I)}\mathbf{N}_n^{(1)}\right]\ ,\label{E1}\\
    \mathbf{H}_1^{\rm (I)}&=&-\imath\frac{k_1}{\omega\mu_1}\sum_{n=-\infty}^{\infty}E_n\left[c_n^{\rm(I)}\mathbf{M}_n^{(1)}+d_n^{\rm(I)}\mathbf{N}_n^{(1)}\right]\label{H1}\
    ;\\
   \mathbf{E}_1^{\rm (II)}&=&-\imath\sum_{n=-\infty}^{\infty}E_n\left[d_n^{\rm(II)}\mathbf{M}_n^{(1)}+c_n^{\rm(II)}\mathbf{N}_n^{(1)}\right]\ ,\label{E2}\\
   \mathbf{H}_1^{\rm (II)}&=&-\frac{k_1}{\omega\mu_1}\sum_{n=-\infty}^{\infty}E_n\left[c_n^{\rm(II)}\mathbf{M}_n^{(1)}+d_n^{\rm(II)}\mathbf{N}_n^{(1)}\right]\label{H2}\
   ,
\end{eqnarray}
where $E_n={rE_0(-\imath)^n}/{\rho_1}$, with $\rho_1=kr(m^2-\cos^2\zeta)^{1/2}$, and the index $(1)$ indicates the Bessel function $J_n(\rho_1)$ to generate the cylindrical harmonics \cite{bohren}.
The quantity $m=({{\mu_1\epsilon_1}/{\mu\epsilon}})^{1/2}$ is the relative refraction index between the cylinder and the surrounding medium, and $k_1=mk$ is the wavenumber inside the cylinder.

To simplify the expressions of the internal ($c_n$, $d_n$) and scattering ($a_n$, $b_n$) coefficients, consider the functions:
\begin{eqnarray}
    \mathcal{A}_n&=&\imath \xi\left[\xi J_n(\xi)J_n'(\eta)\frac{m}{\widetilde{m}}-\eta J_n'(\xi)J_n(\eta)\right]\ ,\label{f1}\\
    \mathcal{B}_n&=&\xi\left[m\widetilde{m}\xi J_n(\xi)J_n'(\eta)-\eta J_n'(\xi)J_n(\eta)\right]\ ,\label{f2}\\
    \mathcal{C}_n&=&n\cos\zeta\ \eta J_n(\xi)J_n(\eta)\left(\frac{\xi^2}{\eta^2}-1\right)\ ,\label{f3}\\
    \mathcal{D}_n&=&n\cos\zeta\ \eta H_n^{(1)}(\xi)J_n(\eta)\left(\frac{\xi^2}{\eta^2}-1\right)\ ,\label{f4}\\
    \mathcal{V}_n&=&\xi\left[m\widetilde{m}\xi H_n^{(1)}(\xi)J_n'(\eta)-\eta H_n'^{(1)}(\xi)J_n(\eta)\right]\ ,\label{f5}\\
    \mathcal{W}_n&=&\imath \xi\left[\eta H_n'^{(1)}(\xi)J_n(\eta)-\xi H_n^{(1)}(\xi)J_n'(\eta)\frac{m}{\widetilde{m}}\right]\ ,\label{f6}
\end{eqnarray}
where $\xi=x\sin\zeta$, $\eta=x({m^2-\cos^2\zeta})^{1/2},\ x=ka$ is the size parameter, $H_n^{(1)}=J_n+\imath Y_n$ is the Hankel function, and $\widetilde{m}=({\mu\epsilon_1/\mu_1\epsilon})^{1/2}$ is the relative impedance between the cylinder and the surrounding medium.
These functions (\ref{f1})--(\ref{f6}) are analogue to those presented in \cite{bohren}, and they are the same for $\mu=\mu_1$ (nonmagnetic approach).

For the TM mode, the boundary conditions provides a set of four
linear equations connecting the coefficients $a_{n}^{(\rm I)}$, $b_{n}^{(\rm I)}$, $\ c_{n}^{(\rm I)}$, and $d_{n}^{(\rm I)}$ \cite{wait,bohren,hulst}.
Solving the system of equations, we obtain:
\begin{eqnarray}
    a_{n}^{(\rm I)}&=&\frac{\mathcal{C}_n\mathcal{V}_n-\mathcal{B}_n\mathcal{D}_n}{\mathcal{V}_n\mathcal{W}_n+\imath \mathcal{D}_n^2}\ ,\label{an-tm}\\
    b_{n}^{(\rm I)}&=&\frac{\mathcal{B}_n\mathcal{W}_n+\imath \mathcal{C}_n\mathcal{D}_n}{\mathcal{V}_n\mathcal{W}_n+\imath \mathcal{D}_n^2}\ ,\label{bn-tm}\\
    c_{n}^{(\rm I)}&=&\frac{-2\imath m\xi\mathcal{W}_n}{\pi\left[\mathcal{V}_n\mathcal{W}_n+\imath \mathcal{D}_n^2\right]}\ ,\label{cn-tm}\\
    d_{n}^{(\rm I)}&=&\frac{-2m\xi\mathcal{D}_n}{\pi\widetilde{m}\left[\mathcal{V}_n\mathcal{W}_n+\imath\mathcal{D}_n^2\right]}\ .\label{dn-tm}
\end{eqnarray}
Similarly, for the TE mode, we have:
\begin{eqnarray}
    a_{n}^{(\rm II)}&=&-\frac{\mathcal{A}_n\mathcal{V}_n-\imath \mathcal{C}_n\mathcal{D}_n}{\mathcal{V}_n\mathcal{W}_n+\imath \mathcal{D}_n^2}\ ,\label{an-te}\\
    b_{n}^{(\rm II)}&=&-\imath \frac{\mathcal{C}_n\mathcal{W}_n+\mathcal{A}_n\mathcal{D}_n}{\mathcal{V}_n\mathcal{W}_n+\imath \mathcal{D}_n^2}\ ,\label{bn-te}\\
    c_{n}^{(\rm II)}&=&\frac{-2 m\xi\mathcal{D}_n}{\pi\left[\mathcal{V}_n\mathcal{W}_n+\imath\mathcal{D}_n^2\right]}\ ,\label{cn-te}\\
    d_{n}^{(\rm II)}&=&\frac{-2m\xi \mathcal{V}_n}{\pi\widetilde{m}\left[\mathcal{V}_n\mathcal{W}_n+\imath \mathcal{D}_n^2\right]}\ ,\label{dn-te}
\end{eqnarray}
where the functions $\mathcal{A}_n,\ \mathcal{B}_n,\ \mathcal{C}_n,\
\mathcal{D}_n,\ \mathcal{V}_n$ and $\mathcal{W}_n$ are defined in Eqs.~(\ref{f1})--(\ref{f6}).

For normal EM wave incidence to the cylinder axis ($\zeta=90^{\rm o}$), we have $a_{n}^{\rm(I)}=b_n^{\rm(II)}=c_n^{\rm(II)}=d_n^{\rm(I)}=0$ and \cite{rayleigh}
\begin{equation*}\begin{split}
    a_n&=a_{n}^{(\rm II)}\big|_{\zeta=90^{\rm o}}=\frac{\widetilde{m}J_n'(x)J_n(mx)-J_n(x)J_n'(mx)}{\widetilde{m}J_n(mx)H_n'^{(1)}(x)-J_n'(mx)H_n^{(1)}(x)}\ ,\\
    b_n&=b_{n}^{(\rm I)}\big|_{\zeta=90^{\rm o}}=\frac{J_n(mx){J_n}'(x)-\widetilde{m}{J_n}'(mx)J_n(x)}{J_n(mx){H_n'^{(1)}}(x)-\widetilde{m}{J_n}'(mx)H_n^{(1)}(x)}\ ,\\
    c_n&=c_{n}^{(\rm I)}\big|_{\zeta=90^{\rm o}}=\frac{2\imath/\pi x}{J_n(mx)H_n'^{(1)}(x)-\widetilde{m}J_n'(mx)H_n^{(1)}(x)}\ ,\\
    d_n&=d_{n}^{(\rm II)}\big|_{\zeta=90^{\rm o}}=\frac{2\imath/\pi x}{\widetilde{m}J_n(mx)H_n'^{(1)}(x)-J_n'(mx)H_n^{(1)}(x)}\
    ,
\end{split}\end{equation*}
where we have used the Wronskian $H_n'^{(1)}(x)J_n(x)-H_n^{(1)}(x)J_n'(x)=2\imath/\pi x$.

\section{Time-averaged internal energy}
\label{internal}

The time-averaged EM energy within a nondispersive finite cylinder with radius $a$ and length $L$ is given by \cite{ruppin,landau}
\begin{equation}\begin{split}
    W(a)=&\int_0^a{\rm d}r\;r\int_0^{2\pi}{\rm d}\phi\int_{-L/2}^{L/2}{\rm d}z\ {\rm Re}\bigg[\frac{\epsilon_1}{4}\big(|E_{1r}|^2+|E_{1\phi}|^2\\
    & +|E_{1z}|^2\big)
    +\frac{\mu_1}{4}\left(|H_{1r}|^2+|H_{1\phi}|^2+|H_{1z}|^2\right)\bigg]\;.\label{ener}
\end{split}\end{equation}
This expression takes $\epsilon_1$ and $\mu_1$ as complex quantities with positive real parts and small imaginary parts compared to the real ones.
In particular, for a cylinder with the same optical properties as the surrounding medium, one has:
\begin{equation}
    W_0=\frac{\pi a^2}{2}\epsilon\left|E_0\right|^2L\;.
\end{equation}
To simplify the analytical expressions and, thereby, the numerical calculations, it is common to use some relations involving the Bessel functions.
Specially, for the average EM energy, \cite{watson} provides two equations in which the integrals associated with product of two cylindrical Bessel functions are performed analytically.
In our notation, for the situation in which there is absorption ($m\not=m^*$), we can define the function
\begin{eqnarray}
    \mathcal{I}_n(\eta)&=&\frac{1}{a^2}\int_0^a{\rm{d}}r\ r{\left|J_n(\rho_1)\right|^2}\nonumber\\
                       &=&2{\rm
                       Re}\left[\frac{\eta^*J_n'(\eta^*)J_n(\eta)}{\eta^2-\eta^{*2}}\right]\;,\label{integral1}
\end{eqnarray}
where $\rho_1(r)=kr(m^2-\cos^2\zeta)^{1/2}$ and $\eta=\rho_1(a)$.
Using the L'Hospital's rule and the recurrence relation $J_n'(\rho)=\pm[nJ_n(\rho)/\rho-J_{n\pm1}(\rho)]$, for real relative refractive index ($m$), Eq.~(\ref{integral1}) can be rewritten as
\begin{eqnarray}
    \mathcal{I}_n(\eta)&=&\frac{1}{a^2}\int_0^a{\rm d}r\;rJ_n^2(\rho_1)\nonumber\\
                        &=&\frac{1}{2}\left[J_n^2(\eta)-J_{n-1}(\eta)J_{n+1}(\eta)\right]\;.\label{integral2}
\end{eqnarray}

In addition, from the recurrence relations $2nJ_n(\rho)=\rho\left[J_{n-1}(\rho)+J_{n+1}(\rho)\right]$ and $2J_n'(\rho)=J_{n-1}(\rho)-J_{n+1}(\rho)$, one can readily show that
\begin{eqnarray}
    &&2\bigg|AJ_n'(\rho)-B\frac{nJ_n(\rho)}{\rho}\bigg|^2+2\left|A\frac{nJ_n(\rho)}{\rho}-BJ_n'(\rho)\right|^2\nonumber\\
                        &&\ =\left|J_{n-1}(\rho)\left(A-B\right)\right|^2+\left|J_{n+1}(\rho)\left(A+B\right)\right|^2\;,\label{bessel1}
\end{eqnarray}
\begin{eqnarray}
    &&2\left|AJ_n'(\rho)+B\frac{nJ_n(\rho)}{\rho}\right|^2+2\left|A\frac{nJ_n(\rho)}{\rho}-BJ_n'(\rho)\right|^2\nonumber\\
    &&\ =\left(|A|^2+|B|^2\right)\left[|J_{n-1}(\rho)|^2+|J_{n+1}(\rho)|^2\right]\nonumber\\
    &&\quad-4{\rm Im}\left(AB^*\right){\rm
    Im}\left[J_{n+1}(\rho)J_{n-1}(\rho^*)\right]\ ,\label{bessel2}
\end{eqnarray}
for any functions $A$ and $B$.
Eqs.~(\ref{bessel1}) and (\ref{bessel2}) are original and they appear in the calculation of the average energy associated with the components $(r,\phi)$ of the EM field at oblique incidence.

Consider the internal fields defined by Eqs.~(\ref{E1})--(\ref{H2}) and take separately each one of the field components in the definition (\ref{ener}).
For the TM polarization, the average EM energy $W_{\rm tot}^{\rm(I)}(a)$ is given by:
\begin{equation}
    W_{\rm tot}^{\rm(I)}=\left[W_{Er}^{\rm(I)}+W_{E\phi}^{\rm(I)}+W_{Ez}^{\rm(I)}\right]+\left[W_{Hr}^{\rm(I)}+W_{H\phi}^{\rm(I)}+W_{Hz}^{\rm(I)}\right]\ ,
\end{equation}
with
\begin{eqnarray}
    {W_{Er}^{\rm (I)}(a)}&=&W_0{\rm Re}\left(m\widetilde{m}\right)\Bigg[\cos^2\zeta\left|\frac{c_0^{(\rm I)}}{m}\right|^2\mathcal{I}_1(\eta)\nonumber\\
                                & &\ +2\sum_{n=1}^{\infty}\int_0^a{\rm d}r\;r\frac{\left|J_n(\rho_1)\right|^2}{a^{2}}\nonumber\\
                                &
                                &\ \times\left|\frac{\cos\zeta}{m}c_n^{(\rm I)}D_n(\rho_1)+d_n^{(\rm I)}\frac{n}{\rho_1}\right|^2\Bigg]\;,\label{Er1}
\end{eqnarray}
\begin{eqnarray}
    {W_{{E}\phi}^{\rm (I)}(a)}&=&2W_0{\rm Re}\left(m\widetilde{m}\right)\sum_{n=1}^{\infty}\int_0^a{\rm d}r\;r\frac{\left|J_n(\rho_1)\right|^2}{a^2}\nonumber\\
    &
    &\ \times\left|\frac{\cos\zeta}{m}c_n^{(\rm I)}\frac{n}{\rho_1}-d_n^{(\rm I)}D_n(\rho_1)\right|^2\;,\label{Ephi1}
\end{eqnarray}
\begin{eqnarray}
    W_{Ez}^{\rm(I)}(a)&=&W_0{\rm
    Re}\left(m\widetilde{m}\right)\left|\frac{\eta}{mx}\right|^2\Bigg[\left|c_0^{\rm(I)}\right|^2\mathcal{I}_{0}(\eta)\nonumber\\
    & &\ +2\sum_{n=1}^{\infty}\left|c_n^{\rm(I)}\right|^2\mathcal{I}_{n}(\eta)\Bigg]\;,\label{Ez1}
\end{eqnarray}
\begin{eqnarray}
    {W_{{H}r}^{\rm I}(a)}&=&2W_0{\rm Re}\left(m\widetilde{m}^*\right)\sum_{n=1}^{\infty}\int_0^a{\rm d}r\;r\frac{\left|J_n(\rho_1)\right|^2}{a^2}\nonumber\\
    &
    &\ \times\left|c_n^{(\rm I)}\frac{n}{\rho_1}-\frac{\cos\zeta}{m}d_n^{(\rm I)}D_n(\rho_1)\right|^2\;,\label{Hr1}
\end{eqnarray}
\begin{eqnarray}
    {W_{H\phi}^{\rm I}(a)}&=&W_0{\rm Re}\left(m\widetilde{m}^*\right)\Bigg[|c_0^{(\rm I)}|^2\mathcal{I}_1(\eta)\nonumber\\
                                & &\ +2\sum_{n=1}^{\infty}\int_0^a{\rm d}r\;r\frac{\left|J_n(\rho_1)\right|^2}{a^2}\nonumber\\
                                & &\ \times\left|c_n^{(\rm I)}D_n(\rho_1)-\frac{\cos\zeta}{m}d_n^{(\rm I)}\frac{n}{\rho_1}\right|^2\Bigg]\;,\label{Hphi1}
\end{eqnarray}
\begin{eqnarray}
    W_{Hz}^{\rm(I)}(a)=2W_0{\rm
    Re}\left(m\widetilde{m}^*\right)\left|\frac{\eta}{mx}\right|^2\sum_{n=1}^{\infty}\left|d_n^{\rm(I)}\right|^2\mathcal{I}_{n}(\eta)\
    ,
\end{eqnarray}
where $W_{Er}=\int{\rm d}rr\int{\rm d}\phi\int{\rm d}z{\rm Re}(\epsilon_1)|E_{1r}|^2/4$, $W_{Hr}=\int{\rm d}rr\int{\rm d}\phi\int{\rm d}z{\rm Re}(\mu_1)|H_{1r}|^2/4$, and so on, and $D_n(\rho_1)=J_n'(\rho_1)/J_n(\rho_1)$.

Because of the integrals in the radial component, observe that Eqs.~(\ref{Er1}), (\ref{Ephi1}), (\ref{Hr1}) and (\ref{Hphi1}) cannot be solved analytically.
However, if one considers the contributions $W_{Er\phi}=(W_{Er}+W_{E\phi})$ and $W_{Hr\phi}=(W_{Hr}+W_{H\phi})$ to the internal energy, the expressions can be simplified by means of Eqs.~(\ref{bessel1}) and (\ref{bessel2}).
Explicitly, using Eq.~(\ref{bessel2}), for $A=c_{n}^{(\rm I)}\cos\zeta/m$ and $B=d_{n}^{(\rm I)}$, it follows from Eqs.~(\ref{Er1}) and (\ref{Ephi1}) that:
\begin{equation}\begin{split}
    W_{Er\phi}^{\rm(I)}(a)&=W_0{\rm
    Re}\left(m\widetilde{m}\right)\Bigg\{{\cos^2\zeta}\left|\frac{c_0^{\rm(I)}}{m}\right|^2\mathcal{I}_{1}(\eta)\\
    &\ +\sum_{n=1}^{\infty}\Bigg[\left({\cos^2\zeta}\left|\frac{c_n^{\rm(I)}}{m}\right|^2+\left|d_n^{\rm(I)}\right|^2\right)\\
    &\ \times\left[\mathcal{I}_{n-1}(\eta)+\mathcal{I}_{n+1}(\eta)\right]\\
    &\ -\frac{4\cos\zeta}{a^2}{\rm
    Im}\left(\frac{
    c_n^{\rm(I)}d_n^{\rm(I)*}}{m}\right)\\
    &\ \times\int_0^a{\rm
    d}r\ r{\rm
    Im}\left[J_{n+1}(\rho_1)J_{n-1}(\rho_1^*)\right]\Bigg]\Bigg\}\;.\label{WE-rphi1}
\end{split}\end{equation}

Also, employing Eq.~(\ref{bessel1}), for $A=c_{n}^{(\rm I)}$ and $B=d_{n}^{(\rm I)}\cos\zeta/m$, we obtain from Eqs.~(\ref{Hr1}) and (\ref{Hphi1}):
\begin{eqnarray}
    W_{Hr\phi}^{\rm(I)}(a)&=&W_0{\rm
    Re}\left(m\widetilde{m}^*\right)\Bigg\{\left|c_0^{\rm(I)}\right|^2\mathcal{I}_{1}(\eta)\nonumber\\
    & &+\sum_{n=1}^{\infty}\bigg[\left|c_n^{\rm(I)}-d_n^{\rm(I)}\frac{\cos\zeta}{m}\right|^2\mathcal{I}_{n-1}(\eta)\nonumber\\
    &
    &+\left|c_n^{\rm(I)}+d_n^{\rm(I)}\frac{\cos\zeta}{m}\right|^2\mathcal{I}_{n+1}(\eta)\bigg]\Bigg\}\
    .\label{WH-rphi1}
\end{eqnarray}
Note that the integral in the last term in Eq.~(\ref{WE-rphi1}) cannot be performed analytically.
However, for normal incidence ($\zeta=90^{\rm o}$: $\mathbf{E}_i||z$), it is clear that $W_{Er\phi}^{\rm(I)}=W_{Hz}^{\rm(I)}=0$, and the average EM energy is $W_{\rm tot}^{\rm(I)}=W_{Ez}^{\rm(I)}+W_{Hr\phi}^{\rm(I)}$.
In this particular case, it can be shown that $\mathcal{I}_{n-1}(y)+\mathcal{I}_{n+1}(y)=4{\rm Re}[yJ_n(y)J_n'(y^*)/(y^2-y^{*2})]$, where $y=mx$, and, thereby,

\begin{eqnarray}
    W_{\rm tot}^{||}&=&
    \frac{2W_0}{x}\sum_{n=-\infty}^{\infty}{\rm
    Re}\bigg\{\left[\frac{J_n(mx)J_n'(m^*x)}{m^2-m^{*2}}\right]\nonumber\\
    &&\ \times\left[m{\rm Re}\left(m\widetilde{m}^*\right)+m^*{\rm
    Re}\left(m\widetilde{m}\right)\right]\bigg\}\left|c_n\right|^2 .\label{W-norm1}
\end{eqnarray}
When $m$ and $\widetilde{m}$ are real quantities, by using L'Hospital's rule, Eq.~(\ref{W-norm1}) takes the simple form
    \begin{eqnarray}
      W_{\rm
      tot}^{||}&=&\frac{W_0\widetilde{m}}{x}\sum_{n=-\infty}^{\infty}\big[J_n(mx)J_n'(mx)+mxJ_n(mx)^2\nonumber\\
      &&\quad-mxJ_{n-1}(mx)J_{n+1}(mx)\big]\left|c_n\right|^2\
      .\label{wreal}
    \end{eqnarray}

For the TE polarization (case II), we obtain similar expressions:

\begin{eqnarray}
{W_{{E}r\phi}^{\rm (II)}(a)}&=&W_{Er}^{\rm II}(a)+W_{E\phi}^{\rm II}(a)\nonumber\\
    &=&W_0{\rm
    Re}\left(m\widetilde{m}\right)\Bigg\{\left|d_0^{\rm(II)}\right|^2\mathcal{I}_{1}(\eta)\nonumber\\
    & &\ +\sum_{n=1}^{\infty}\Bigg[\left|d_n^{\rm(II)}-c_n^{\rm(II)}\frac{\cos\zeta}{m}\right|^2\mathcal{I}_{n-1}(\eta)\nonumber\\
    & &\ +\left|d_n^{\rm(II)}+c_n^{\rm(II)}\frac{\cos\zeta}{m}\right|^2\mathcal{I}_{n+1}(\eta)\Bigg]\Bigg\}\ ,
\end{eqnarray}

\begin{eqnarray}
    W_{Ez}^{\rm(II)}(a)=2W_0{\rm
    Re}\left(m\widetilde{m}\right)\left|\frac{\eta}{mx}\right|^2\sum_{n=1}^{\infty}\left|c_n^{\rm(II)}\right|^2\mathcal{I}_{n}(\eta)\
    ,
\end{eqnarray}

\begin{equation}\begin{split}
    W_{Hr\phi}^{\rm(II)}(a)&=W_0{\rm
    Re}\left(m\widetilde{m}^*\right)\Bigg\{\cos^2\zeta\left|\frac{d_0^{\rm(II)}}{m}\right|^2\mathcal{I}_{1}(\eta)\\
    &\ +\sum_{n=1}^{\infty}\Bigg[\left({\cos^2\zeta}\left|\frac{d_n^{\rm(II)}}{m}\right|^2+\left|c_n^{\rm(II)}\right|^2\right)\\
    &\ \times\left[\mathcal{I}_{n-1}(\eta)+\mathcal{I}_{n+1}(\eta)\right]\\
    &\ -\frac{4\cos\zeta}{a^2}{\rm
    Im}\left(\frac{
    d_n^{\rm(II)}c_n^{\rm(II)*}}{m}\right)\\
    &\ \times\int_0^a{\rm
    d}r\ r{\rm
    Im}\left[J_{n+1}(\rho_1)J_{n-1}(\rho_1^*)\right]\Bigg]\Bigg\}\ .
\end{split}\end{equation}
\begin{eqnarray}
        W_{Hz}^{\rm(II)}(a)&=&W_0{\rm Re}\left(m\widetilde{m}^*\right)\left|\frac{\eta}{mx}\right|^2\Bigg[\left|d_0^{\rm(II)}\right|^2\mathcal{I}_{0}(\eta)\nonumber\\
   & &\ +2\sum_{n=1}^{\infty}\left|d_n^{\rm(II)}\right|^2\mathcal{I}_{n}(\eta)\Bigg]\ .
\end{eqnarray}
For normal incidence ($\zeta=90^{\rm o}$: $\mathbf{E}_i\perp z$), one obtains that $W_{Ez}^{\rm(II)}=W_{Hr\phi}^{\rm(II)}=0$ and, therefore, the average EM energy is given by $W_{\rm tot}^{\rm(II)}=W_{Er\phi}^{\rm(II)}+W_{Hz}^{\rm(II)}$.
Explicitly, we obtain
\begin{eqnarray}
    W_{\rm tot}^{\perp}&=&
    \frac{2W_0}{x}\sum_{n=-\infty}^{\infty}{\rm
    Re}\bigg\{\left[\frac{J_n(mx)J_n'(m^*x)}{m^2-m^{*2}}\right]\nonumber\\
    &&\ \times\left[m^*{\rm Re}\left(m\widetilde{m}^*\right)+m{\rm
    Re}\left(m\widetilde{m}\right)\right]\bigg\}\left|d_n\right|^2 .\label{W-norm2}
\end{eqnarray}
If $m$ and $\widetilde{m}$ are real quantities, Eq.~(\ref{W-norm2}) becomes Eq.~(\ref{wreal}) replacing $c_n$ with $d_n$.

In all equations above, we have used the equalities
${\rm Re}\left(m\widetilde{m}\right)={\rm Re}\left(\epsilon_1\right)/\epsilon$, which is associated with the electric field, and
${\rm Re}\left(m\widetilde{m}^*\right)=\left|k_1/\omega\mu_1\right|^2{\rm Re}\left(\mu_1\right)/\epsilon$, which appears in the magnetic one.
For dispersive cylinders, the expressions for the internal energy must be modified according to the model used to write the functions $\epsilon_1(\omega)$ and $\mu_1(\omega)$ \cite{ruppin,landau}.

\section{Absorption efficiency}
\label{absorption}

The efficiencies in the EM scattering by a non-optically active infinite cylinder are
\begin{eqnarray}
Q_{\rm sca}^{\rm(I)}&=&\frac{2}{x}\left[\left|b_0^{(\rm
I)}\right|^2+2\sum_{n=1}^{\infty}\left(\left|b_{n}^{(\rm I)}\right|^2
+\left|a_{n}^{(\rm I)}\right|^2\right)\right]\ ,\\
Q_{\rm tot}^{\rm(I)}&=&\frac{2}{x}{\rm Re}\left[b_0^{(\rm I)}+2\sum_{n=1}^{\infty}b_{n}^{(\rm I)}\right]\ ,\\
Q_{\rm abs}^{\rm(I)}&=&Q_{\rm tot}^{\rm(I)}-Q_{\rm sca}^{\rm(I)}\ ,\label{eff1}
\end{eqnarray}
where $Q_{\rm tot}^{\rm(I)}$, $Q_{\rm sca}^{\rm(I)}$ and $Q_{\rm abs}^{\rm(I)}$ are the extinction (or total), scattering and absorption efficiencies for the TM polarization, repectively.
Expressions for the TE mode are obtained replacing $a_{n}^{(\rm I)}$ with $b_{n}^{(\rm II)}$ and $b_{n}^{(\rm I)}$ with $a_{n}^{(\rm II)}$ \cite{bohren}.

Using the boundary conditions for the TM mode
\begin{eqnarray}
    \eta J_n(\eta)c_n^{\rm (I)}&=&m\xi \left[J_n(\xi)-H_n^{(1)}(\xi)b_n^{\rm
    (I)}\right]\ ,\label{cc1} \\
    \imath\widetilde{m}\eta J_n(\eta)d_n^{\rm(I)}&=&m\xi
    H_n^{(1)}(\xi)a_n^{\rm (I)}\ ,\label{dd1}
\end{eqnarray}
and the definitions for the magnetic internal coefficients $c_n^{\rm(I)}$ and $d_n^{\rm(I)}$, given by Eqs.~(\ref{cn-tm}) and (\ref{dn-tm}), we obtain
\begin{equation}\begin{split}
    Q_{\rm abs}^{\rm(I)}&=\frac{2}{x}\sum_{n=-\infty}^{\infty}{\rm Re}\Bigg\{\left|c_n^{\rm(I)}\right|^2
    \frac{\imath\pi\widetilde{m}}{2m^*}\eta J_n(\eta^*)J_n'(\eta)+\left|\frac{\widetilde{m}d_n^{\rm(I)}}{m}\right|^2\\
    &\ \times\left[\frac{\pi\eta^2J_n(\eta^*)\left(\xi^2-\eta^{*2}\right)\mathcal{W}_n}
    {2\xi^2\eta^{*2}H_n^{(1)}(\xi)\left(\eta^2-\xi^2\right)}-\left|\frac{\eta J_n(\eta)}{H_n^{(1)}(\xi)}\right|^2\right]\Bigg\}\
    ,\label{abs01}
\end{split}\end{equation}
where $\mathcal{W}_n$ is defined in Eq.~(\ref{f6}).
In the same manner, using the boundary conditions for the TE mode
    \begin{eqnarray}
        \eta J_n(\eta)c_n^{\rm(II)}&=&\imath m\xi
        H_n^{(1)}(\xi)b_n^{\rm(II)}\ ,\label{cc2}\\
        \widetilde{m}\eta
        J_n(\eta)d_n^{\rm(II)}&=&m\xi\left[J_n(\xi)-H_n^{(1)}(\xi)a_n^{\rm(II)}\right]\
        ,\label{dd2}
    \end{eqnarray}
and the coefficients $c_n^{\rm(II)}$ and $d_n^{\rm(II)}$, given by Eqs.~(\ref{cn-te}) and (\ref{dn-te}), we obtain

\begin{equation}\begin{split}
    Q_{\rm abs}^{\rm(II)}&=\frac{2}{x}\sum_{n=-\infty}^{\infty}{\rm Re}\Bigg\{\left|\frac{\widetilde{m}d_n^{\rm(II)}}{m}\right|^2
    \frac{\imath\pi m}{2\widetilde{m}}\eta^* J_n(\eta^*)J_n'(\eta)+\left|\frac{c_n^{\rm(II)}}{m}\right|^2\\
    &\ \times\left[\frac{\imath\pi\eta^2J_n(\eta^*)\left(\xi^2-\eta^{*2}\right)\mathcal{V}_n}
    {2\xi^2\eta^{*}H_n^{(1)}(\xi)\left(\xi^2-\eta^2\right)}-\left|\frac{\eta J_n(\eta)}{H_n^{(1)}(\xi)}\right|^2\right]\Bigg\}\
    ,\label{abs02}
\end{split}\end{equation}
where $\mathcal{V}_n$ is defined in Eq.~(\ref{f5}).

If we consider in the last term of Eq.~(\ref{abs01}) the approximation $\eta^2\approx\eta^{*2}$, which means $m^2\approx m^{*2}$ (low absorption), it can readily be shown that
\begin{eqnarray}
    Q_{\rm abs}^{\rm(I)}&\approx&\frac{\pi}{x}\sum_{n=-\infty}^{\infty}\Bigg\{\left|{c_n^{\rm(I)}}\right|^2{\rm
    Im}\left[\frac{\widetilde{m}^*}{m}\eta
    J_n(\eta)J_n'(\eta^*)\right]\nonumber\\
    &&\quad+\left|d_n^{\rm(I)}\right|^2{\rm
    Im}\left[\frac{\widetilde{m}}{m}\eta J_n(\eta)
    J_n'(\eta^*)\right]\Bigg\}\ ,\label{Qabs}
\end{eqnarray}
where we have used the Wronskian $H_n'^{(1)}(\xi)J_n(\xi)-H_n^{(1)}(\xi)J_n'(\xi)=2\imath/\pi\xi$.
An analogous expression is obtained for the TE polarization by replacing the index (I) with (II) in Eq.~(\ref{Qabs}).

It is important to emphasize that in Eq.~(\ref{Qabs}) only the terms that vanish for the normal incidence are approximated by using $\eta^2\approx\eta^{*2}$.
Therefore, when $\zeta=90^{\rm o}$, it follows that

\begin{eqnarray}
    Q_{\rm abs}^{||}&=&{\pi}\sum_{n=-\infty}^{\infty}{\rm
    Im}\left[{\widetilde{m}^*}J_n(mx)J_n'(m^*x)\right]\left|{c_n}\right|^2\ ,\label{Qabs1}\\
    Q_{\rm abs}^{\perp}&=&{\pi}\sum_{n=-\infty}^{\infty}{\rm
    Im}\left[{\widetilde{m}}J_n(mx)
    J_n'(m^*x)\right]\left|d_n\right|^2\ ,\label{Qabs2}
\end{eqnarray}
which are exact expressions for the parallel and perpendicular absorption efficiencies expanded in terms of the magnetic internal coefficients $c_n$ and $d_n$.
In the following, we show an expected interrelation between the average EM energy within a cylinder and its optical-absorption efficiency, provided that ${\rm Re}(m)\gg {\rm Im}(m)$ and ${\rm Re}(\widetilde{m})\gg{\rm Im}(\widetilde{m})$ (low absorption limit).

\section{Numerical calculations}
\label{numerical}

\begin{figure}[htb]
\centering
\includegraphics[angle=0, width=1\linewidth]{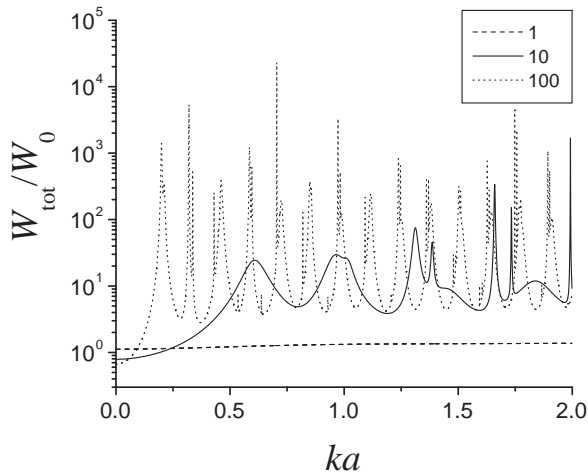}
\caption{Normalized EM energy $W_{\rm tot}^{\rm(I)}/W_0$ within a magnetic ($\mu_1/\mu=10$, 100) and a nonmagnetic ($\mu_1/\mu=1$) cylinder with $\epsilon_1/\epsilon=(1.334 + 1.5\times 10^{-9})^2$. Only the TM polarization is shown, with $\zeta=60^{\rm o}$.}
\label{fig1}
\end{figure}

\begin{figure}[htb]
\centering
\includegraphics[angle=0, width=1\linewidth]{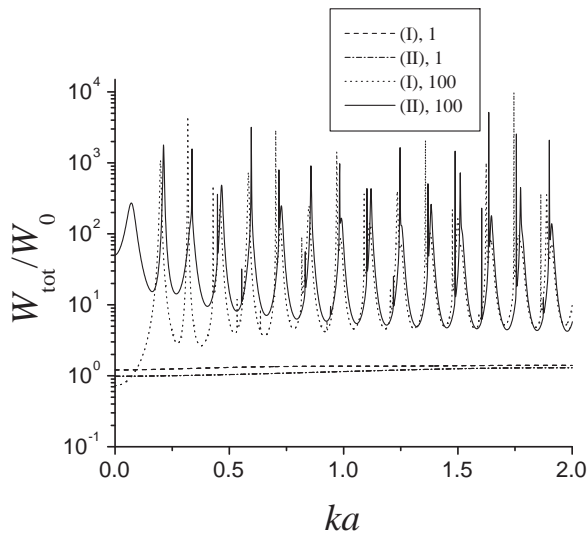}
\caption{Comparison between the normalized EM energy $W_{\rm tot}/W_0$ within a magnetic ($\mu_1/\mu=100$) and a nonmagnetic ($\mu_1/\mu=1$) cylinder with $\epsilon_1/\epsilon=1.4161$. The parallel and perpendicular polarizations are indicated by (I) and (II), respectively.}
\label{fig2}
\end{figure}

\begin{figure}[htb]
\centering
\includegraphics[angle=0, width=1\linewidth]{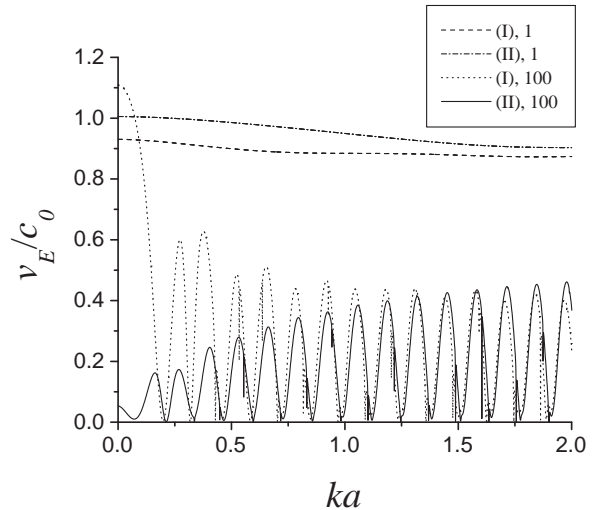}
\caption{Comparison between the normalized energy-transport velocity $v_E/c_0$ in a medium containing magnetic ($\mu_1/\mu=100$) and nonmagnetic ($\mu_1/\mu=1$) cylinders with $\epsilon_1/\epsilon=1.4161$ and volume fraction $f=0.36$. The parallel and perpendicular polarizations are indicated by (I) and (II), respectively.}
\label{fig3}
\end{figure}

\begin{figure}[htb]
\centering
\includegraphics[angle=0, width=1\linewidth]{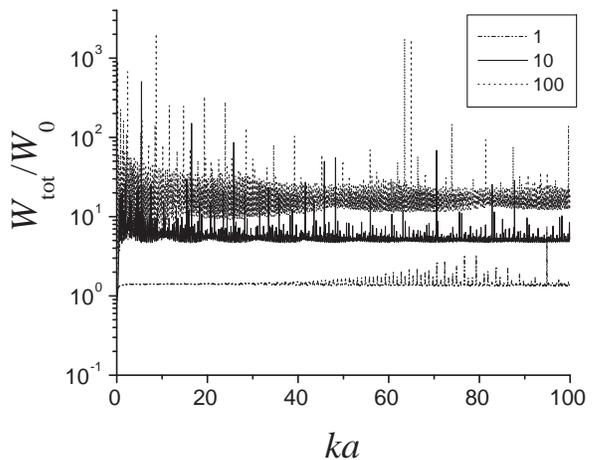}
\caption{Normalized EM energy $W_{\rm tot}^{\rm(I)}/W_0$ within a magnetic ($\mu_1/\mu=10$, 100) and a nonmagnetic ($\mu_1/\mu=1$) cylinder with $\epsilon_1/\epsilon=1.4161$, parallel polarization.}
\label{fig4}
\end{figure}

\begin{figure}[htb]
\centering
\includegraphics[angle=0, width=1\linewidth]{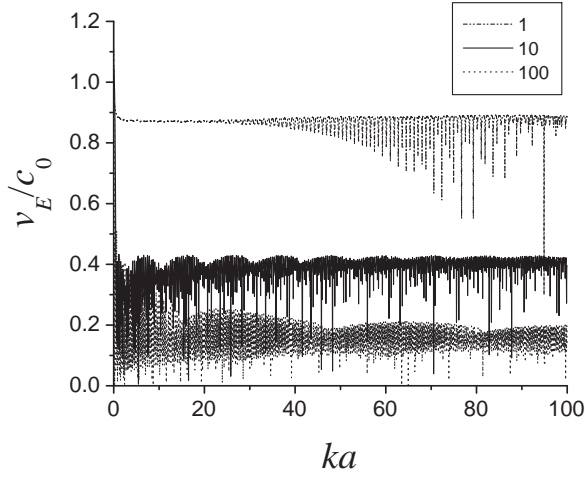}
\caption{Normalized energy-transport velocity $v_E/c_0$ in a medium containing magnetic ($\mu_1/\mu=10$, 100) and nonmagnetic ($\mu_1/\mu=1$) cylinders with $\epsilon_1/\epsilon=1.4161$ and volume fraction $f=0.36$, parallel polarization.}
\label{fig5}
\end{figure}

\begin{figure}[htb]
\centering
\includegraphics[angle=0, width=1\linewidth]{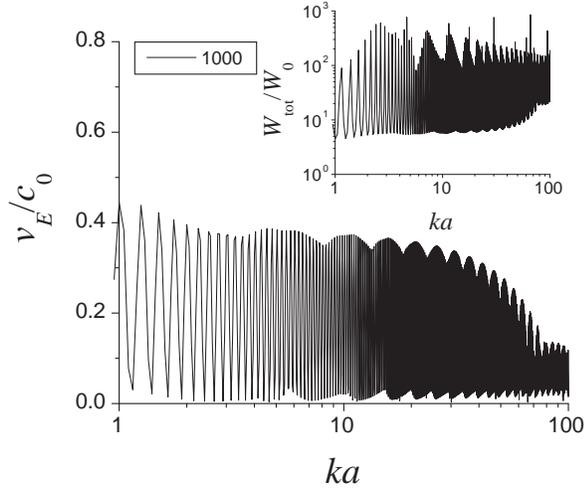}
\caption{Normalized energy-transport velocity $v_E/c_0$ in a medium containing magnetic cylinders ($\mu_1/\mu=1000$), parallel polarization.}
\label{fig6}
\end{figure}

Here we present some numerical results from the exact expressions of the time-averaged EM energy within a magnetic cylinder.
All numerical calculations have been performed by programs written for the free software for scientific computation $Scilab\ 5.1.1$.
As an upper limit $N$ for the truncated series $\sum_{n=1}^{N}$, we employ the expression $N=\max(n_c,|m|x)+(101.0+x)^{1/2}$, with $n_c=x+4.05x^{1/3}+2$ \cite{barber}.
The modification added in $N$ in which we take the value $\max(n_c,|m|x)$ instead of only $n_c$ is introduced to give more accurate sums even for large values of $\mu_1/\mu$ at small values of $x$ (Figs.~\ref{fig1}, \ref{fig2} and \ref{fig3}).
The exceptions are the Figs.~\ref{fig4}, \ref{fig5} and \ref{fig6}, where we replace $\max(n_c,|m|x)$ with $n_c$.

Fig.~\ref{fig1} shows a comparison between the magnetic and nonmagnetic approach for the TM mode with $\zeta=60^{\rm o}$ (oblique incidence).
The quantities are calculated in the interval $0<x<2$, with $\delta x=10^{-3}$.
We use in this calculation the same value of $\epsilon_1/\epsilon$ of \cite{alexandre-new} for a magnetic sphere with a small imaginary part added.
The result achieved in Fig.~\ref{fig1} is quite similar to the one we have obtained for EM scattering by a magnetic sphere \cite{tiago}.
In both cases, the average internal energy is much larger than that one related to a scatterer with the same optical properties as the surrounding medium even for $x<1$.

The series of sharp peaks in Fig.~\ref{fig1} as a function of the size parameter are well-known and are generally referred to as morphology-dependent resonances (MDR) \cite{michenko}.
For the average internal energy, they are ascribed to the resonances of the far-field scattering coefficients \cite{owen,petr}, which are related to the internal coefficients by Eqs.~(\ref{cc1}), (\ref{dd1}), (\ref{cc2}) and (\ref{dd2}).
Physically, these large values of the normalized internal energy can be explained by the enhancement of the extinction efficiency $Q_{\rm tot}$, whose sharp peaks occur at the corresponding size parameters in which the large values of $W_{\rm tot}/W_0$ take place \cite{ruppin}.

\subsection{Weak absorption}

Similarly to the case studied in \cite{bott,tiago}, it can be shown that for a cylindrical scatterer with weak absorption (wa) there is a relation between the average EM energy and the optical-absorption efficiency.
For sake of simplicity, consider Eq.~(\ref{W-norm1}) and assume that $m_i\ll m_r$ and $\widetilde{m}_i\ll \widetilde{m}_r$, where $m=m_r+\imath m_i$, $\widetilde{m}=\widetilde{m}_r+\imath\widetilde{m}_i$ are the complex relative refraction and impedance indexes, respectively.
Approximating $(m^2-m^{*2})\approx4\imath m_r m_i$, ${\rm Re}(m\widetilde{m}^*)\approx m_r\widetilde{m}_r$ and ${\rm Re}(m\widetilde{m})\approx m_r\widetilde{m}_r$, we have:
\begin{eqnarray}
    \frac{W_{\rm
    tot}^{\rm (I)}}{W_0}\approx\sum_{n=-\infty}^{\infty}\frac{m_r}{m_ix}{\rm
    Im}\left[\widetilde{m}_rJ_n(mx)J_n'(m^*x)\right]\left|c_n\right|^2\
    .\label{W-approx1}
\end{eqnarray}
Comparing Eqs.~(\ref{Qabs1}) and (\ref{W-approx1}), it results that
\begin{eqnarray}
    \frac{W_{\rm tot}^{\rm (wa)}}{W_0}\approx\left(\frac{m_r}{\pi
    m_ix}\right)Q_{\rm abs}\ .\label{WQ}
\end{eqnarray}
The relation (\ref{WQ}) for cylindrical scatterers, which to the best of our knowledge has not been determined so far, holds for both TM and TE modes and it does not depend on the incidence angle $(90^{\rm o}-\zeta)$.

An analogous result has been obtained for the EM scattering by a dielectric sphere \cite{bott} and a magnetic one \cite{tiago}.
Due to the system symmetry, the constant which associates $W_{\rm tot}/W_0$ with $Q_{\rm abs}$ for a homogeneous magnetic sphere is not the same as that one for an infinitely long cylinder:
$W_{\rm tot}^{\rm (sph)}/W_0\approx3m_rQ_{\rm abs}^{\rm (sph)}/(8m_ix)$.
In the special case of the infinite cylinder with optical properties similar to the surrounding medium, i.e., $m_r\approx1$ and $W_{\rm tot}^{\rm(wa)}\approx W_0$, it follows that: $Q_{\rm abs}\approx\pi m_i x$.
This last result is in agreement with \cite{hulst}. For the sphere, one has $Q_{\rm abs}^{\rm(sph)}\approx 8m_ix/3$ \cite{bott,tiago,hulst}.

As a curiosity, comparing the EM scattering by spheres with infinite cylinders, both in the weak absorption regime ($m_i\ll m_r$), we can rewrite Eq.~(\ref{WQ}) as
\begin{eqnarray}
    \frac{W_{\rm tot}^{\rm (wa)}}{W_0}&\approx&\frac{a\sigma_{\rm
    g}}{2V}\frac{m_r}{x}\frac{Q_{\rm abs}}{m_i}\nonumber\\
    &=&\frac{m_r}{2kV}\frac{\sigma_{\rm abs}}{m_i}\
    ,\label{especial}
\end{eqnarray}
where $V$ is the volume of the scatterer, $k$ is the wavenumber of the incident EM wave, and $\sigma_{\rm g}$ and $\sigma_{\rm abs}=Q_{\rm abs}\sigma_{\rm g}$ are the geometrical and the absorption cross sections, respectively.
Note that, although we explicitly consider two particular geometries in this derivation (a sphere of radius $a$ and a segment $L$ of an infinite cylinder of radius $a$), Eq.~(\ref{especial}) does not depend on the shape of the scatterer.
Of course, Eq.~(\ref{especial}) must be further investigated to verify if its universality is valid or not.

\subsection{Energy-transport velocity}

Van Tigglen et al. \cite{bart} have shown that, for simple 3D dielectric scatterers, the energy-transport velocity $v_E$ is related to the energy-enhancement factor $W_{\rm tot}/W_0$ by the expression
$v_E = c_0/\left[1+f(W_{\rm tot}/W_0-1)\right]$, where $c_0$ is the wave velocity in the host medium ($\epsilon,\mu$) and $f$ is the volume fraction occupied by the scatterers.
Ruppin~\cite{ruppin} has successfully used this expression for $v_E$ as an application of the average energy stored inside an infinite dielectric cylinder.
As it has been reported in \cite{ruppin}, this simple model to calculate the transport velocity in 2D medium reproduces well the results of Busch et al. \cite{busch}, obtained in a different context of the low-density approximation of the Bethe-Salpeter equation.

Here, we extend the use of this expression for $v_E$ to the calculation of the energy-transport velocity in a 2D disordered magnetic medium ($\zeta=90^{\rm o}$).
We consider a random collection of parallel isotropic cylinders with a packing fraction $f=0.36$, which is the same used in the experiments with nonmagnetic scatterers TiO$_2$ \cite{bart}.
Specially, we assume the scatterers are magnetic and have negligible losses, i.e., $(\epsilon_1,\mu_1)$ are real quantities.
This last assumption can be achieved in soft ferrites, which present large values of $\mu_1/\mu$ with low magnetic losses at microwave frequencies typically below 100 MHz \cite{sigalas}.

The quantities in Figs.~\ref{fig2} and \ref{fig3} are calculated in the interval $0<x<2$, with $\delta x=10^{-3}$.
The resonance peaks in the EM internal energy (Fig.~\ref{fig2}), as expected, provide small values of the energy-transport velocity (Fig.\ref{fig3}) \cite{ruppin,bart}.
Because of the magnetism, the transport velocity vanish even for cylinders with radius much smaller than the wavelength (Rayleigh size region).
Vanishing of energy-transport velocity for $x<1$ in a 3D disordered magnetic medium has been reported in \cite{alexandre-vanish}.
Qualitatively, this means that the EM wave spends a long time (dwell time) inside the scatterers, leading to a decrease in $v_E$ and, thereby, in the diffusion coefficient $D=v_E\ell^{\star}/3$, where $\ell^{\star}$ is the transport mean free path \cite{alexandre-vanish,bart,busch}.
This strong decrease in the transport velocity, and consequently in the diffusion coefficient, is related to the single scatterer resonances and leads to electromagnetic wave localization \cite{bart,busch}.
The decrease of $v_E$ in a 2D disordered magnetic medium can be observed in Figs.~\ref{fig3} (for $0<x< 2$) and \ref{fig5} (for $0<x< 100$).

Notice that the energy-transport velocity plotted as a function of the size parameter shows an oscillatory behavior.
Because of the differences in the average EM energy between the parallel and perpendicular polarizations (Fig.~\ref{fig2}), the behavior of the normalized transport velocity is also different for both polarizations.
Indeed, they show opposite oscillatory tendencies for small size parameters: while the oscillation amplitude of $v_E/c_0$ in the parallel mode is reduced with increasing $x$, in the perpendicular mode it is increased.
This can be clearly observed in Fig.~\ref{fig3}.

In the interval $1<x<100$, with $\delta x=0.05$, we show in Fig.~\ref{fig6} the profile of $v_E/c_0$ (parallel mode) for $\mu_1/\mu=1000$.
Although the number of shaper drops in this size parameter region, for $\delta x< 0.05$, is much larger than it is represented here, one can observe the global oscillatory behavior of $v_E$ as a functions of $x$.
In a particular configuration for $x=0.01$, \cite{alexandre-vanish} has studied the quantity $v_E/c_0$ as a function of the relative magnetic permeability $\mu_1/\mu$.

\section{Conclusion}

The time-averaged EM energy inside an irradiated magnetic cylinder, for a general case of oblique incidence, has analytically been calculated for the TM and TE modes.
We have shown that, similar to the dielectric \cite{bott} and magnetic \cite{tiago} spheres with low absorption, the optical-absorption efficiency associated with a magnetic cylinder, taken to be weakly absorptive, is related to its internal energy-enhancement factor.
Indeed, this particular result, when compared to that one from single Mie scattering, suggests a more general relation which does not depend on the shape of the scattering center.
If its universality is valid, it can be applied to calculate in a simple way the energy-transport velocity in a disordered weakly absorptive media, provided only the volume and the absorption cross section of the particles embedded in the medium.
Finally, we have used the normalized internal EM energy to determine the energy-transport velocity for 2D disordered magnetic medium.
We have shown, in particular, the vanish of the transport velocity even in the Rayleigh size region.

\section*{Acknowledgements}

The authors acknowledge the support of the Brazilian agencies National Counsel of Technological and Scientific Development (CNPq) (303990/2007-4 and 476862/2007-8) and the State of São Paulo Research Foundation (FAPESP) (2008/02069-0).

\appendix

\section{Approximations for the normal incidence }
\label{appendice}

Below we consider the scattering coefficients in the small magnetic particle limit.
Two other limits obtained are the Rayleigh and ferromagnetic ones.
The scattering amplitudes are then calculated to address polarization in the Rayleigh limit.

\subsection{Small-particle limit}

If $x\ll1$, we can approximate the values of the scattering coefficients $a_n$ and $b_n$ using the limiting cases of the cylindrical Bessel functions \cite{bohren}.
For simplicity, consider the cylinder is normally illuminated ($\zeta=90^{\rm o}$).
The values of these coefficients for $n=0$ and $n=1$ are:
\begin{equation}\begin{split}
    a_0&\approx\frac{\imath\pi D_0(mx)}{8\widetilde{m}}x\left(4-{x^2}\right)+\frac{\imath\pi}{32}x^2\left(8-{x^2}\right)\
    ,\\
    b_0&\approx\frac{\imath\pi\widetilde{m}{D_0(mx)}}{8}x\left(4-{x^2}\right)+\frac{\imath\pi}{32}x^2\left(8-{x^2}\right)\
    ,\\
    a_1&\approx\frac{-\imath\pi
    x^2\left[\widetilde{m}\left(8-3x^2\right)-D_1(mx)x\left(8-x^2\right)\right]}{32\left[D_1(mx)x+\widetilde{m}\right]}\
    ,\\
    b_1&\approx\frac{-\imath\pi
    x^2\left[8-3x^2-\widetilde{m}D_1(mx)x\left(8-x^2\right)\right]}{32\left[\widetilde{m}D_1(mx)x+1\right]}\
    .\label{thin}
\end{split}\end{equation}

\subsection{Rayleigh and ferromagnetic limits}

In the Rayleigh approximation, $x\ll1$ and $|m|x\ll1$, which leads to
$D_0(mx)\approx-{mx}/{2}-{m^3x^3}/{16}$ and $D_1(mx)\approx{1}/{mx}-{mx}/{4}$,
we obtain:
\begin{equation*}\begin{split}
    a_0&\approx-\frac{\imath\pi}{4}\left(\frac{m}{\widetilde{m}}-1\right)x^2-\frac{\imath\pi}{32}\left(1+\frac{m^3}{\widetilde{m}}-2\frac{m}{\widetilde{m}}\right)x^4\
    ,\\
    b_0&\approx-\frac{\imath\pi}{4}\left(m{\widetilde{m}}-1\right)x^2-\frac{\imath\pi}{32}\left(1+{m^3}{\widetilde{m}}-{2m}{\widetilde{m}}\right)x^4\
    ,\\
    a_1&\approx-\frac{\imath\pi}{4}\left(\frac{m\widetilde{m}-1}{m\widetilde{m}+1}\right)x^2-\frac{\imath\pi}{32}\left[\frac{2m^2-3m\widetilde{m}+1}{m\widetilde{m}+1}\right]x^4\
    ,\\
    b_1&\approx-\frac{\imath\pi}{4}\left(\frac{m/\widetilde{m}-1}{m/\widetilde{m}+1}\right)x^2-\frac{\imath\pi}{32}\left[\frac{2m^2-3m/\widetilde{m}+1}{m/\widetilde{m}+1}\right]x^4\
    .\label{ray}
\end{split}\end{equation*}
Note that, in the magnetic approach, the terms of order $x^2$ in the coefficients $a_0$ and $b_1$ do not vanish.
The expressions presented in \cite{bohren} for the nonmagnetic scattering are recovered when $m=\widetilde{m}$.

The ferromagnetic limit is also derived from the small-particle limit.
However, we must consider both $x\ll1$ and $|m|x\gg1$.
For large arguments, the logarithmic derivative function can be written as
\begin{equation}
 D_n(mx)\approx-\frac{1}{mx}-\tan\left[mx-(2n+1)\pi/4\right]\
 ,\label{ap}
\end{equation}
and the ferromagnetic approximation is obtained by substitution of $D_0(mx)$ and $D_1(mx)$ calculated from Eq.~(\ref{ap}) into the far-field scattering coefficients of small-particle limit, Eqs.~(\ref{thin}).

\subsection{Polarization}

For a normally irradiated cylinder, the degree of polarization of the scattered light is $P={T_{11}}/{T_{12}}$, where $T_{11}=(|T_1|^2-|T_2|^2)/2$, $T_{12}=(|T_1|^2+|T_2|^2)/2$,
and the amplitude functions are $T_1=b_{0}+2\sum_{n=1}^{\infty}b_{n}\cos(n\Theta)$ and $T_2=a_{0}+2\sum_{n=1}^{\infty}a_{n}\cos(n\Theta)$, with $\Theta=180^{\rm o}-\phi$ \cite{bohren}.

Using Eqs.~(\ref{thin}) in the Rayleigh limit, one can write the amplitude functions $T_1$ and $T_2$ in terms of order
$x^2$:
\begin{equation*}\begin{split}
    T_1&\approx-\frac{\imath\pi}{4}(m\widetilde{m}-1)x^2-\frac{\imath\pi}{2}\left(\frac{m/\widetilde{m}-1}{m/\widetilde{m}+1}\right)x^2\cos\Theta\
        ,\\
    T_2&\approx-\frac{\imath\pi}{4}\left(\frac{m}{\widetilde{m}}-1\right)x^2-\frac{\imath\pi}{2}\left(\frac{m\widetilde{m}-1}{m\widetilde{m}+1}\right)x^2\cos\Theta\
        .
\end{split}\end{equation*}
At $\Theta=90^{\rm o}$, the value of degree of polarization $P$ of the scattered light is not identically 1 for
a magnetic cylinder:
\begin{eqnarray}
    P=\frac{\left|m\widetilde{m}-1\right|^2-\left|m/\widetilde{m}-1\right|^2}{\left|m\widetilde{m}-1\right|^2+\left|m/\widetilde{m}-1\right|^2}\
    .
\end{eqnarray}
Therefore, similarly to the EM scattering by a magnetic sphere
\cite{kerkermag}, the radiant intensity scattered by a magnetic cylinder is not symmetrical about
$90^{\rm o}$.


\newpage
\begin{thebibliography}{99}

\bibitem{sigalas}
M. M. Sigalas, C. M. Soukoulis, R. Biswas, and K. M. Ho,
``Effect of the magnetic permeability on photonic band gaps,''
Phys. Rev. B {\bf 56}, 959-962 (1997).

\bibitem{ping}
P. Chen, R. X. Wu, J. Xu, A. M. Jiang, and X. Y. Ji,
``Effects of magnetic anisotropy on the stop band of ferromagnetic electromagnetic band gap materials,''
J. Phys. Condens. Matter {\bf 19}, 106205 (2007).

\bibitem{lin2}
Z. F. Lin, and S. T. Chui,
``Manipulating electromagnetic radiation with magnetic photonic crystals,''
Opts. Lett. {\bf 32}, 2288-2290 (2007).

\bibitem{liu}
S. Liu, J. Du, Z. Lin, R. X. Wu, and S. T. Chui,
``Formation of robust and completely tunable resonant photonic band gaps,''
Phys. Rev. B {\bf 78}, 155101 (2008).

\bibitem{lin}
Z. F. Lin, and S. T. Chui,
``Electromagnetic scattering by optically anisotropic magnetic particles,''
Phys. Rev. E {\bf 69}, 056614 (2004).

\bibitem{tarento}
R.-J. Tarento, K.-H. Bennemann, P. Joyes, and J. Van de Walle,
``Mie scattering of magnetic spheres,''
Phys. Rev. E {\bf 69}, 026606 (2004).


\bibitem{alexandre-vanish}
F. A. Pinheiro, A. S. Martinez, and L. C. Sampaio,
``Vanishing of energy transport and diffusion constant of electromagnetic waves in disordered magnetic media,''
Phys. Rev. Lett. {\bf 85}, 5563-5566 (2000).

\bibitem{alexandre-multi}
F. A. Pinheiro, A. S. Martinez, and L. C. Sampaio,
``Multiple scattering of electromagnetic waves in disordered magnetic media: localization parameter, energy transport velocity and diffusion constant,''
Braz. J. Phys. {\bf 31}, 65-70 (2001).

\bibitem{kerkermag}
M. Kerker, D. S. Wang, and C. L. Giles,
``Electromagnetic scattering by magnetic spheres,''
J. Opt. Soc. Am. {\bf 73}, 765-767 (1983).

\bibitem{alexandre-new}
F. A. Pinheiro, A. S. Martinez, and L. C. Sampaio, ``New effects
in light scattering in disordered media and coherent
backscattering cone: system of magnetic particles,'' Phys. Rev.
Lett.  {\bf 84}, 1435-1438 (2000).

\bibitem{alexandre-elec}
F. A. Pinheiro, A. S. Martinez, and L. C. Sampaio,
``Electromagnetic scattering by small magnetic particles,'' J.
Magnet. Mag. Mat. {\bf 226-230},  1951-1953 (2001).


\bibitem{tiago}
T. J. Arruda, and A. S. Martinez, ``Electromagnetic energy within
magnetic spheres,'' J. Opt. Soc. Am. A {\bf 27}, 992-1001 (2010).

\bibitem{rayleigh} Lord Rayleigh,
``The dispersal of light by a dielectric cylinder,''
 Philos. Mag. {\bf 36}, 365-376 (1918).

\bibitem{stratton}
J. A. Stratton,
\emph{Electromagnetic Theory}
(McGraw-Hill, 1941).


\bibitem{wait} J. R. Wait,
``Scattering of a plane wave from a circular dielectric cylinder at oblique incidence'', Can. J. Phys. {\bf 33}, 189-195 (1955).

\bibitem{lind} A. C. Lind, and J. M. Greenberg,
``Electromagnetic scattering by obliquely oriented cylinders'',
J. Appl. Phys. {\bf 37} 3195-3203 (1966).

\bibitem{ruppin}
R. Ruppin,
``Electromagnetic energy inside an irradiated cylinder,''
J. Opt. Soc. Am. A {\bf 15}, 1891-1895 (1998).


\bibitem{bart}
B. A. van Tiggelen, A. Lagendijk, M. P. van Albada, and A. Tip,
``Speed of light in random media,''
Phys. Rev. B {\bf 45}, 12233-12243 (1992).

\bibitem{bohren}
C. F. Bohren, and D. R. Huffman, \emph{Absorption and Scattering
of Light by Small Particles} (Wiley, 1983).

\bibitem{bott}
A. Bott, and W. Zdunkowski,
``Electromagnetic energy within dielectric spheres,''
J. Opt. Soc. Am. A {\bf 4}, 1361-1365 (1987).


\bibitem{barber}
P. W. Barber,
\emph{Light Scattering by Particles: Computational Methods}
(World Scientific, 1990).


\bibitem{hulst}
H. C. van de Hulst,
\emph{Light Scattering by Small Particles}
(Dover, 1980).


\bibitem{kerker}
M. Kerker,
\emph{The Scattering of Light and other Electromagnetic Radiation}
(Academic, 1969).


\bibitem{landau}
L. D. Landau, E. M. Lifshitz, and L. P. Pitaevskii.
\emph{Electrodynamics of Continuous Media} (Pergamon, 1984).


\bibitem{watson}
G. N. Watson,
\emph{A Treatise on the Theory of Bessel Functions}
(Cambridge Mathematical Library, 1958).


\bibitem{michenko}
M. I. Mishchenko, and A. A. Lacis,
``Manifestations of morphology-dependent resonances in Mie scattering matrices,''
Appl. Math. and Comp. {\bf 116}, 167-179 (2000).

\bibitem{owen}
J. F. Owen, R. K. Chang, and P. W. Barber,
``Internal electric field distributions of a dielectric cylinder at resonance wavelengths,''
Opt. Lett. {\bf 6}, 540-542 (1981).


\bibitem{petr}
P. Chylek, J. D. Pendleton, and R. G. Pinnick,
``Internal and near-surface scattered field of a spherical particle at
resonant conditions,''
Appl. Opts. {\bf 24}, 3940-3942 (1985).



\bibitem{busch}
K. Busch, C. M. Soukoulis, and E. N. Economou,
``Transport velocity in two-dimensional random media,''
Phys. Rev. B {\bf 52}, 10834-10840 (1995).

\end{thebibliography}
\end{document}